\documentstyle[11pt,newpasp,twoside,epsf]{article}
\def\edcomment#1{\iffalse\marginpar{\raggedright\sl#1\/}\else\relax\fi}
\reversemarginpar

\def\te{\Delta t_{\rm Earth}}

\def\tf{\Delta t_{\rm frustrated}}
\def\tb{\Delta t_{\rm biogenesis}}
\def\tl{\Delta t_{\rm life}}

\def\tn{\Delta t(N)}

\begin{document}
\title{What can rapid terrestrial biogenesis tell us about life in the universe?}

\author{Charles H. Lineweaver \& Tamara M. Davis}
\affil{School of Physics, University of New South Wales and the\\
 Australian Centre for Astrobiolgy, Sydney, Australia\\
charley@bat.phys.unsw.edu.au}

\begin{abstract}
It is sometimes asserted that the rapidity of biogenesis on Earth suggests that life
is common in the Universe.
We critically examine the assumptions inherent in this argument.
Using a lottery model for biogenesis in the Universe,
we convert the observational constraints on the rapidity of 
biogenesis on Earth into constraints on the probability of biogenesis 
on other terrestrial planets.
For example, if terrestrial biogenesis took less than 200 Myr 
(and we assume that it could have taken 1 billion years) then we find 
the probability of biogenesis on terrestrial planets older than 
$\sim 1$ Gyr,  is $> 36\%$ at the $95\%$ confidence level.
However, there are assumptions and selection effects that complicate
this result:
although we correct the analysis for the fact that
biogenesis is a prerequisite for our existence,
our result depends on the plausible assumption
that {\bf{\it rapid}} biogenesis is 
not such a prerequisite.
\end{abstract}
Our existence on Earth can tell us little about 
how common life is in the Universe or about the probability 
of biogenesis on a terrestrial planet because, 
even if this probability were infinitesimally small and 
there were only one life-harboring planet in the Universe 
we would, of necessity, find ourselves on that planet.
However, the rapidity with which life appeared on Earth gives
us more information.
We find ourselves in the group of planets on which biogenesis has necessarily 
occurred -- we have of necessity won the biogenetic lottery some time 
in the past. And we also find that biogenesis has occurred rapidly -- we won 
soon after the tickets went on sale.
From radioactivity to lotteries, more probable things happen more rapidly.
If life were rare it would be unlikely that biogenesis would have occurred 
as rapidly as it seems to have occurred on Earth.

During and immediately following the Earth's 
formation there was a period  without life ($\tf$),  
followed by a period during which life evolved ($\tb$), 
followed by a period during which life has been present ($\tl$).
The sum of these intervals adds up to the age of the Earth (Fig. 1):
\begin{equation}
\tf + \tb + \tl = \te.
\end{equation}
Inserting observational estimates for these durations (Lineweaver \& Davis 2002, hereafter LD) yields 
\begin{equation}
0.5 \pm 0.4 + \tb + 4.0^{+0.4}_{-0.2} = 4.566 \pm 0.002 \;{\rm {\mbox Gyr}}
\end{equation}
or
\begin{equation}
\tb= 0.1^{+0.5}_{-0.1}\; {\rm {\mbox Gyr.}}
\end{equation}
\clearpage
\begin{figure}[h,t!]
\plotone{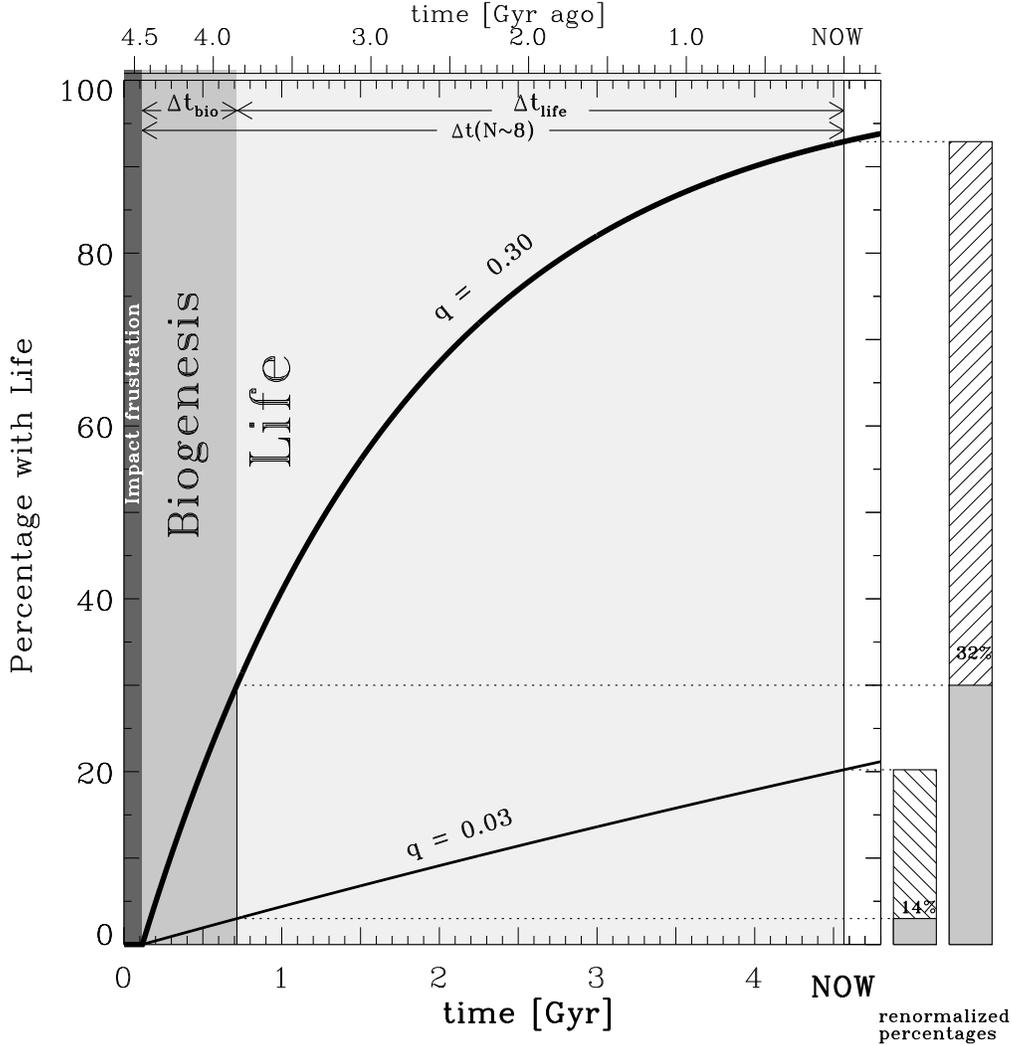}
\caption{We divide the history of the Earth into three epochs: 
Impact frustration, Biogenesis and Life.
Consider a group of terrestrial planets with approximately the same probability of biogenesis
`$q$' as Earth. Suppose $q= 0.30$. At their
formation, none of these planets had life. As time passed, life arose on
more and more of them (thick line)
After $\tb$, $30\%$ will have life (that is how $q=0.30$ is defined).
After $4.566$ Gyr, $93\%$ will have life ($7\%$ still will not).
If $q$ is high ($0.30$, thick line) a large fraction ($32\%$) 
of the planets that have evolved life within $4.566$ Gyr of formation,
have life that evolved rapidly -- within $\tb$ --
on their planets. If $q$ is low ($0.03$, thin line) then a 
smaller fraction ($14\%$) will have life that evolved rapidly.
These different percentages illustrate the principle that
a single observation of rapid terrestrial biogenesis
is more likely to be the result of high $q$ and allows us to compute the 
relative likelihood of $q$.}
\end{figure}
\clearpage
\begin{figure}
\plotfiddle{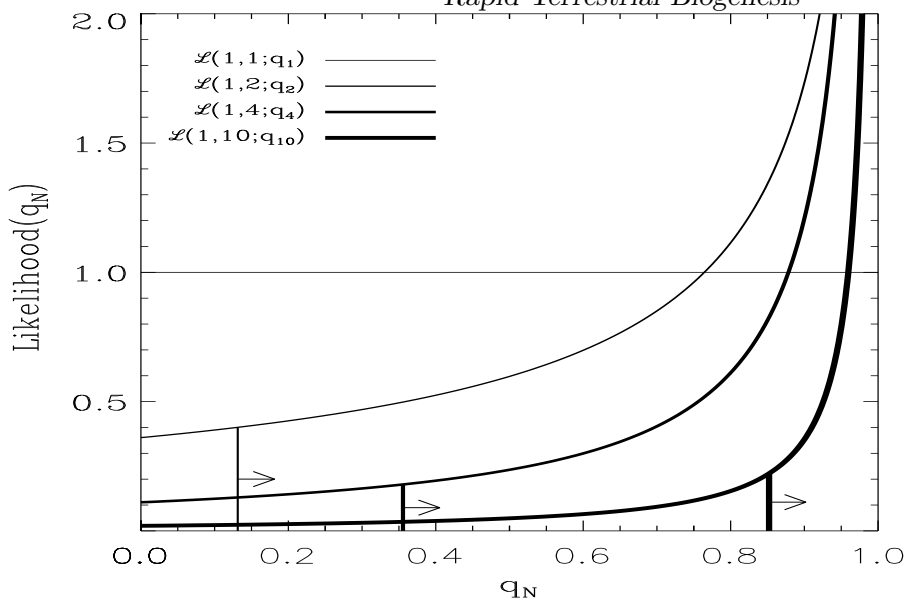}{6.6cm}{0}{70}{43}{-220}{-50}      
\caption{Likelihood of $q_{N}$.
Here we generalize inferences about $q$ (the probability of 
biogenesis within $\tb$)  to inferences about $q_{N}$ 
(the probability of biogenesis within an arbitrary time, $\tn= N \times \tb$).  
The $95\%$ lower limit on $q_{N}$ increases dramatically as $N$ increases, 
constraining $q_{N}$ to be close to $1$.
Specifically, we only need to be able to assume that 
biogenesis could have been twice as long as $\tb$ (i.e., $N \ge 2$)
to have interesting constraints:
$q_{2} > 0.13$, $q_{4} > 0.36$ and $q_{10} > 0.85$ at the $95\%$ confidence level.
The ability to extract a useful constraint even if $N$ is low reduces
the influence of the non-observability-of-recent-biogenesis selection effect. 
However, estimates of how large $N$ is are problematic. 
If $N=1$ we can say nothing meaningful 
about $q_{1}\: (=q)$. See LD for details.}
\end{figure}

\noindent Thus we take $600$ Myr as a crude estimate of the 
upper limit for the time it took life to appear on Earth.
This is the basis for the statement that biogenesis occurred
rapidly. 
%
We convert this constraint into a constraint on the probability of biogenesis
$q$ using a lottery model and some simple standard statistics (see LD for details).
For this conversion we need to include the fact that 
the evolution of an observer takes some time.
How long it takes observing equipment to evolve, is difficult to say.  
On Earth it took $\sim 4$ Gyr. 
Thus, there is a selection effect favouring 
biogenesis to happen a few billion years before the
present regardless of whether it happened rapidly.
It is not a selection effect for rapid biogenesis since the 
longer it took us to evolve to a point when we could measure the 
age of the Earth, the older the Earth could become.
For example, if biogenesis took $1$ Gyr longer than it actually did, 
we would currently find the age of the Earth to be 
$5.566$ Gyr ($=4.566 + 1$) old.

Any effect that makes rapid biogenesis a 
prerequisite for life would undermine our inferences for $q$.
For example, although it is usually assumed that the heavy bombardment
inhibited biogenesis, energetic impacts may have set up large
chemical and thermal disequilibria that play some crucial role in biogenesis.
If true, the timescale of biogenesis would be linked to the timescale of
the exponential decay of bombardment and biogenesis would (if it occurred
at all) be necessarily rapid; most extant life in the Universe would 
have rapid biogenesis and little could be inferred about the absolute value 
of $q$ from our sample of one.

Carter (1983) has pointed out that the timescale for the evolution of intelligence on the Earth  
($\sim 4$ Gyr) is comparable to the main sequence lifetime of the Sun ($\sim 10$ Gyr).  Under the 
assumption that these two timescales are independent, he argues that this would be unlikely to 
be observed unless the average timescale for the evolution of intelligence on a terrestrial planet 
is much longer than the main sequence lifetime of the host star (see Livio (1999) for an objection 
to the idea that these two timescales are independent).
Carter's argument is strengthened by recent models of the terrestrial biosphere that indicate that the gradual 
increase of solar luminosity will make Earth uninhabitable in 1 or 2 billion years -- several billion years 
before the Sun leaves the main sequence (Caldeira and Kasting 1992). 

Our analysis is similar in style to Carter's, however we are concerned with 
the appearance of the earliest life forms, not the appearance of  
intelligent life.
Subject to the caveats raised by LD and Livio (1999), the implications 
of our analysis and Carter's are consistent and complementary: the appearance of life on terrestrial planets may be 
common but the appearance of intelligent life may be rare.
Hanson (2002) 
applies Carter's reasoning about the emergence of intelligence 
(which took about 4 billion years) to the emergence of life (which may have taken 100 million years or less)
and warns that the Earth may not be a random member of the planets with life -- life had to evolve quickly to
leave lots of time for intelligence to evolve.  However, in the regime where
$\tb$ is much less than the 1-2 billion years remaining for life on Earth (Caldeira and Kasting 1992)
this argument becomes less compelling.

\begin{references}
\reference{Caldeira, K. and Kasting, J.F. (1992) The life span of the biosphere revisited. {\it Nature} 360, 721-723.}
\reference{Carter, B. (1983) The Anthropic Principle and its Implications for Biological Evolution. In  {\it Proc.  R. Soc Discussion
Meeting on the Constants of Physics}, edited by W.H. McRea and M.J. Rees, R. Soc., London and in 
{\it Philos. Trans. R. Soc. London}, A310, 347-355.}
\reference{Hanson, B. (2002) working paper on line at http://hanson.gmu.edu/vita.html
 Must Early Life be Easy? The Rhythm of Major Evolutionary Transitions.}
\reference{Lineweaver, C.H. and Davis, T.M. (2002)  
Does the rapid appearance of life on Earth suggest that life is common in the Universe? {\it Astrobiology}
in press, astro-ph/0205014.}
\reference{Livio, M. (1999) How rare are extraterrestrial civilizations and 
when did they emerge? {\it Astrophys. J.} 511, 429-431.}
\end{references}
\end{document}